# The Schrödinger functional in QCD

Stefan Sint

Max-Planck-Institut für Physik
Werner-Heisenberg-Institut
Föhringer Ring 6, 80805 Munich, Germany

The Schrödinger functional in Wilson's lattice QCD leads to a sensible classical continuum theory which can be taken as starting point for a perturbative analysis. In dimensional regularization, the saddle point expansion of the Schrödinger functional is performed to one-loop order of perturbation theory. The divergences are partly cancelled by the usual coupling constant and quark mass renormalization. An additional divergence can be absorbed in a multiplicative renormalization of the quark boundary fields. The corresponding boundary counterterm being a local polynomial in the fields we confirm the general expectation expressed by Symanzik [1].

## 1. INTRODUCTION

During the last two years, a systematic study of the continuum limit in the pure SU(2) and SU(3) lattice gauge theories has been carried out [2,3]. The success of this analysis is partly due to the use of the Schrödinger functional (SF) as an observable in numerical simulations. In order to include the quark degrees of freedom, one is therefore lead to define and investigate the SF in QCD.

The SF [4] is the euclidean propagation kernel for going from a field configuration at time $x_0 = 0$ to another configuration at euclidean time $x_0 = T$. From the euclidean point of view one has to do with a quantum field theory on a space-time manifold with boundaries, the initial and final field configurations being the boundary values of the euclidean fields.

Perturbative renormalizability of a field theory is usually established using power counting theorems in momentum space. In presence of a boundary, these are no longer applicable and one has to worry about the universality of the continuum limit even in perturbation theory. At this point it is useful to recall Symanzik's work on the Schrödinger representation in field theory [1]. Based on explicit calculations in the $\lambda\phi^4$ theory, he expressed the expectation that the SF in any renormalizable field theory is again renormalizable, provided one includes a finite number of boundary counterterms. These are local polynomials in the fields and derivatives thereof of dimension three or less, integrated over the hyper planes at $x_0 = 0$ and $x_0 = T$.

In the following I briefly recall the way the SF is defined on the lattice and then report on a one-loop evaluation of the SF in dimensional regularization. It will be assumed that the reader is familiar with the pure gauge theory framework of ref.[4], so that only the additional features will be discussed.

## 2. CLASSICAL CONTINUUM THEORY FROM THE LATTICE

On the lattice, the SF is naturally defined as the integral kernel of the $M$th power of the transfer matrix, where $M = T/a$ denotes the number of lattice points in the euclidean time direction ($a$ is the lattice spacing). Going through this procedure with the well-known transfer matrix for Wilson quarks [5], one obtains the following boundary conditions for the quark fields [1] ($P_\pm = \frac{1}{2}(1 \pm \gamma_0)$),

$$P_+\psi|_{x_0=0} = \rho_+, \qquad P_-\psi|_{x_0=T} = \rho'_-,$$
$$\bar\psi P_-|_{x_0=0} = \bar\rho_-, \qquad \bar\psi P_+|_{x_0=T} = \bar\rho'_+. \qquad (1)$$

Further details concerning the lattice functional can be found in ref.[7]. There it has also been noted that the boundary terms in the continuum

---
[1] For the approach with staggered fermions cf. ref.[6]

limit of the fermionic lattice action,

$$\begin{aligned} S_f[A,\bar\psi,\psi] & = \int_0^T \mathrm{d}x_0 \int_0^L \mathrm{d}^3\mathbf{x}\,\bar\psi\{\slashed{D}+m_q\}\psi \\ & \quad - \int_0^L \mathrm{d}^3\mathbf{x}\,[\bar\psi P_-\psi]_{x_0=0} \\ & \quad - \int_0^L \mathrm{d}^3\mathbf{x}\,[\bar\psi P_+\psi]_{x_0=T}, \end{aligned} \quad (2)$$

can be understood without reference to the lattice, once the boundary conditions (1) are known. It suffices to assume parity invariance of the action [2] and the existence of smooth classical solutions $\psi_{cl}$ and $\bar\psi_{cl}$ to the field equations. Writing the quark fields as classical fields plus fluctuations, one obtains

$$S_f[A,\bar\psi_{cl}+\bar v,\psi_{cl}+v] = S_f[A,\bar\psi_{cl},\psi_{cl}] + S_f[A,\bar v,v], \quad (3)$$

with a well-defined Dirac operator acting on the fluctuation fields. Its propagator $S(x,x')$ is also well-defined and can be used to construct the classical solutions, viz

$$\psi_{cl}(x) = \int_0^L \mathrm{d}^3\mathbf{x}'\,[S(x;0,\mathbf{x}')\rho_+(\mathbf{x}') + S(x;T,\mathbf{x}')\rho'_-(\mathbf{x}')], \quad (4)$$

and similarly for $\bar\psi_{cl}$.

## 3. FORMAL CONTINUUM APPROACH

The classical continuum theory is taken as the starting point for a formal definition of the SF in the continuum,

$$\mathcal{Z} = \sum_{n=-\infty}^{\infty} \int \mathrm{D}[\psi]\mathrm{D}[\bar\psi]\mathrm{D}[A]\,\mathrm{e}^{-S[A,\bar\psi,\psi]}, \quad (5)$$

where $\mathcal{Z} = \mathcal{Z}[\bar\rho'_+,\rho'_-,C';\bar\rho_-,\rho_+,C]$ is a functional of the boundary fields. The sum over winding numbers $n$ on top of the functional integrals corresponds to the projection on the gauge invariant subspace, as explained in detail in ref.[4].

At small values of the gauge coupling $g_0$ the dominant contributions to the functional integral are expected from small neighborhoods around the absolute minima of the action. It is assumed here that the minimal action configuration $B_\mu(x)$ is unique up to gauge transformations and occurs in the winding number $n=0$ sector. The gauge group is then given as the group of transformations which leave the boundary gauge fields $C$ and $C'$ intact. If the latter are irreducible (i.e. $C = C^\Lambda$ implies $\Lambda \propto 1$, for any gauge function $\Lambda(\mathbf{x})$), the gauge group $\mathcal{G}$ consists of those transformations $\Omega$ which satisfy [4]

$$\Omega(x) = \begin{cases} z_m & \text{at } x_0 = 0, \\ 1 & \text{at } x_0 = T, \end{cases} \quad (6)$$

where $z_m = \exp(i2\pi m/N)$, $m = 0,\ldots,N-1$. The gauge group $\mathcal{G}$ thus consists of $N$ disconnected components and the gauge orbit of any gauge potential will decompose accordingly. In fact, any saddle point of the pure gauge theory has an $N$-fold degeneracy which is lifted when quarks with non-vanishing quark boundary fields are included.

A nice method to deal with this degeneracy starts with the observation that a gauge transformation (6) in the functional integral leads to

$$\begin{aligned} & \mathcal{Z}[\bar\rho'_+,\rho'_-,C';\bar\rho_-,\rho_+,C] \\ & \quad = \mathcal{Z}[\bar\rho'_+,\rho'_-,C';\bar\rho_- z_m^{-1},z_m\rho_+,C]. \end{aligned} \quad (7)$$

We may thus average over all $m$

$$\mathcal{Z} = \frac{1}{N}\sum_{m=0}^{N-1} \mathcal{Z}[\bar\rho'_+,\rho'_-,C';\bar\rho_- z_m^{-1},z_m\rho_+,C]. \quad (8)$$

This summation can be regarded as part of the functional integral, if the boundary values $z_m$ of the gauge transformation are considered additional "dynamical" variables of the action, transforming as $z_m \to z_{l+m}$ under a gauge transformation (6) with boundary value $z_l$. The achievement is that the enlarged action and the enlarged measure are separately invariant under the action of the gauge group. One may now decompose the gauge field $A = B + g_0 q$ and apply the gauge fixing procedure in the same way as in the pure gauge theory (cf. ref. [4]).

## 4. ONE-LOOP RESULTS

The perturbative evaluation being the same for each value of $m$, one may carry out the summa-tion over $m$ at the very end and define an "effective action" $\Gamma$ for the $m = 0$ component of the gauge fixed functional,

$$\Gamma[B, \bar{\psi}_{cl}, \psi_{cl}] \stackrel{\text{def}}{=} -\ln \mathcal{Z}^{(m=0)}. \tag{9}$$

---

[2] The vacuum angle $\theta$ [8,9] is taken to vanish.

The arguments of $\Gamma$ are the background gauge field and the classical quark fields of sect.2. $\Gamma$ is invariant under *arbitrary* gauge transformations of its arguments and has an asymptotic expansion in the bare gauge coupling constant,

$$\Gamma = g_0^{-2}\,\Gamma_0 + \Gamma_1 + g_0^2\,\Gamma_2 + \ldots. \tag{10}$$

The first coefficients $\Gamma_0$ and $\Gamma_1$ contain the classical action of the background fields, i.e.

$$\Gamma_0[B] = g_0^2\, S_{\text{g}}[B], \tag{11}$$
$$\Gamma_1[B, \bar{\psi}_{cl}, \psi_{cl}] = S_f[B, \bar{\psi}_{cl}, \psi_{cl}] + \ldots. \tag{12}$$

Technical details of the one-loop computation of $\Gamma$ are given elsewhere [10]. Here we quote the results for the poles in $\varepsilon = (4-D)/2$,

$$\Gamma_1 \stackrel{\varepsilon \to 0}{=} -\frac{1}{3\varepsilon}\frac{11N - 2n_f}{16\pi^2}\Gamma_0 + \text{O}(1), \tag{13}$$

$$\Gamma_2 \stackrel{\varepsilon \to 0}{=} \frac{3C_{\text{F}}}{16\pi^2\varepsilon}\left[m_q \int \text{d}^4x\,\bar{\psi}_{cl}(x)\psi_{cl}(x) - S_f[B, \bar{\psi}_{cl}, \psi_{cl}]\right] + \text{O}(1). \tag{14}$$

Here, $n_f$ denotes the number of flavours and $C_{\text{F}} = (N^2 - 1)/2N$ is a constant. The pole multiplying $\Gamma_0$ and the one proportional to the quark mass $m_q$ are eliminated by the usual QCD renormalizations of the coupling constant and the quark mass [11–15]. The interesting effect of the boundary is the pole which multiplies the classical quark action. Using the fact that $\bar{\psi}_{cl}$ and $\psi_{cl}$ are solutions of the classical field equations, one infers that the additional boundary counterterm is of the same form as the boundary terms already present in the classical quark action (2). If one introduces the new renormalization constant $Z_b = 1 + g_{\text{MS}}^2 Z_b^{(1)}(\varepsilon) + \ldots$, and sets

$$Z_b^{(1)}(\varepsilon) = \frac{3C_{\text{F}}}{16\pi^2\varepsilon}, \tag{15}$$

one may conclude that $\Gamma[B, Z_b^{-1/2}\bar{\psi}_{cl}, Z_b^{-1/2}\psi_{cl}]$ is finite to order $g_{\text{MS}}^2$. Recalling eq.(4) and introducing the renormalized quark boundary fields through

$$\rho_+ = Z_b^{1/2}\rho_+^R, \qquad \rho'_- = Z_b^{1/2}\rho'^R_-,$$
$$\bar{\rho}_- = Z_b^{1/2}\bar{\rho}^R_-, \qquad \bar{\rho}'_+ = Z_b^{1/2}\bar{\rho}'^R_+, \tag{16}$$

one may conclude that $\mathcal{Z}[\bar{\rho}'^R_+, \rho'^R_-, C'; \bar{\rho}^R_-, \rho^R_+, C]$ is finite at one-loop order.

Finally, the summation over $m$ is seen to enforce the physical picture that only "colourless", i.e. $SU(N)$ singlet states are propagated by the SF. In perturbation theory, the gauge invariance of the SF is also reflected by the fact that the renormalization constants are independent of the gauge fixing parameter. In particular, $Z_b$ has *nothing* to do with the (gauge dependent) wave function renormalization of the quark field.

In conclusion, the reported one-loop result gives confidence that the SF indeed is a universal quantity in QCD.